\documentclass[prl,twocolumn,showpacs,preprintnumbers,amsmath,amssymb]{revtex4}

\usepackage{graphicx}
\usepackage{dcolumn}
\usepackage{bm}

\begin{document}

\newcommand*{\cm}{cm$^{-1}$\,}
\newcommand*{\Tc}{T$_c$\,}


\title{Origin of the Spin density wave instability in AFe$_2$As$_2$ (A=Ba, Sr) as revealed by optical spectroscopy}

\author{W. Z. Hu}
\author{J. Dong}
\author{G. Li}
\author{Z. Li}
\author{P. Zheng}
\author{G. F. Chen}
\author{J. L. Luo}
\author{N. L. Wang}

\affiliation{Beijing National Laboratory for Condensed Matter
Physics, Institute of Physics, Chinese Academy of Sciences,
Beijing 100190, China}


\begin{abstract}

We performed optical spectroscopy measurement on single crystals
of BaFe$_2$As$_2$ and SrFe$_2$As$_2$, the parent compounds of FeAs
based superconductors. Both are found to be quite metallic with
fairly large plasma frequencies at high temperature. Upon entering
the spin-density-wave (SDW) state, formation of partial energy
gaps was clearly observed with the presence of surprisingly two
different energy scales. A large part of the Drude component was
removed by the gapping of Fermi surfaces (FS). Meanwhile, the
carrier scattering rate was even more dramatically reduced. We
elaborate that the SDW instability is more likely to be driven by
the FS nesting of itinerant electrons rather than a local-exchange
mechanism.
\end{abstract}



\maketitle

The recent discovery of superconductivity with transition
temperature T$_c$ above 50 K in RFeAsO$_{1-x}$F$_x$ (R=La, Ce, Sm,
Pr, Nd, etc) has created tremendous interests in the scientific
community.\cite{Kamihara08,Chen1,XHChen,Ren1}. Those compounds
crystallize in a tetragonal ZrCuSiAs-type structure, which
consists of alternate stacking of edge-sharing Fe$_2$As$_2$
tetrahedral layers and R$_2$O$_2$ tetrahedral layers along c-axis.
The parent compound LaFeAsO itself is not superconducting but
shows strong anomalies near 150 K in resistivity, magnetic
susceptibility, specific heat, etc. Based on experimental
observations and first principle calculations, it is suggested
that the ground state is a spin-density-wave (SDW) ordered state
with a stripe-type (or collinear) spin configuration.\cite{Dong}
The predicted magnetic structure was confirmed by subsequent
neutron diffraction experiment, although the neutron data
indicated that a subtle structural distortion occurs first near
150 K, and the SDW long range order establishes at a slightly
lower temperature.\cite{Cruz} With fluorine doping, the SDW order
is suppressed and superconductivity emerges.\cite{Dong,Chen1} The
very closeness of the superconducting phase to the SDW instability
suggests that the magnetic fluctuations play a key role in the
superconducting pairing mechanism.

Investigating the origin of the antiferromagnetic (AFM) SDW
instability in the parent compound is an essential step towards
understanding the mechanism of superconductivity in doped systems.
The stripe-type AFM order was first suggested to result from the
nesting between the hole and electron Fermi surfaces (FS) of
itinerant electrons.\cite{Dong} Alternatively it was proposed that
the superexchange interaction mediated through the off-plane As
atom plays a key role in the spin configuration
formation.\cite{Yildirim,Si,Ma,Fang,Xu,Wu} A stripe-type AFM would
arise when the next nearest neighbor exchange becomes larger than
half of the nearest neighbor exchange interaction. Whether an
itinerant picture or a local superexchange mechanism should be
taken as a starting approach becomes one of the most important
issues for those systems.\cite{Mazin}

Very recently, it is found that the ThCr$_2$Si$_2$-type ternary
iron arsenide BaFe$_2$As$_2$, which contains identical
edge-sharing Fe$_2$As$_2$ tetrahedral layers as in LaFeAsO,
exhibits a similar SDW instability at 140 K.\cite{Rotter1} It is
therefore suggested that BaFe$_2$As$_2$ could serve as a new
parent compound for ternary iron arsenide superconductors. Shortly
after that, the superconductivity with T$_c$=38 K was found in
K-doped BaFe$_2$As$_2$, which was suggested to be a hole-doped
iron arsenide superconductor.\cite{Rotter2} The SDW instability
was also found in SrFe$_2$As$_2$ with a higher transition
temperature near 200K,\cite{Krellner,Chen2} while K-doping again
introduces the superconductivity with T$_c$$\sim$38 K.\cite{Chen2}
For both parent compounds, the same stripe-type AFM order was
confirmed by neutron experiments.\cite{Huang,Zhao} A great
advantage of those ternary iron arsenide compounds is that it is
much easier to grow large size single crystals.\cite{Ni}

In this letter we present optical study on both BaFe$_2$As$_2$ and
SrFe$_2$As$_2$ single crystals. We find the undoped compounds of
Fe-pnictides are quite metallic with rather high plasma
frequencies, $\omega_p\geq$ 1.5 eV. Upon entering into the SDW
state, formation of energy gaps was clearly observed.
Surprisingly, the optical measurement revealed two distinct energy
scales in the gapped state. Associated with the gapping of the FS,
a large part of Drude component is removed, meanwhile the carrier
scattering rate shows even steeper reduction. Beyond the character
energy for SDW gap, another spectral suppression in R($\omega$)
which covers a much higher energy scale is present even above SDW
transition temperature. The physical implications of those results
were discussed. Our study favors an itinerant electron approach
for the driving mechanism of SDW instability.

\begin{figure*}
\includegraphics[width=7in]{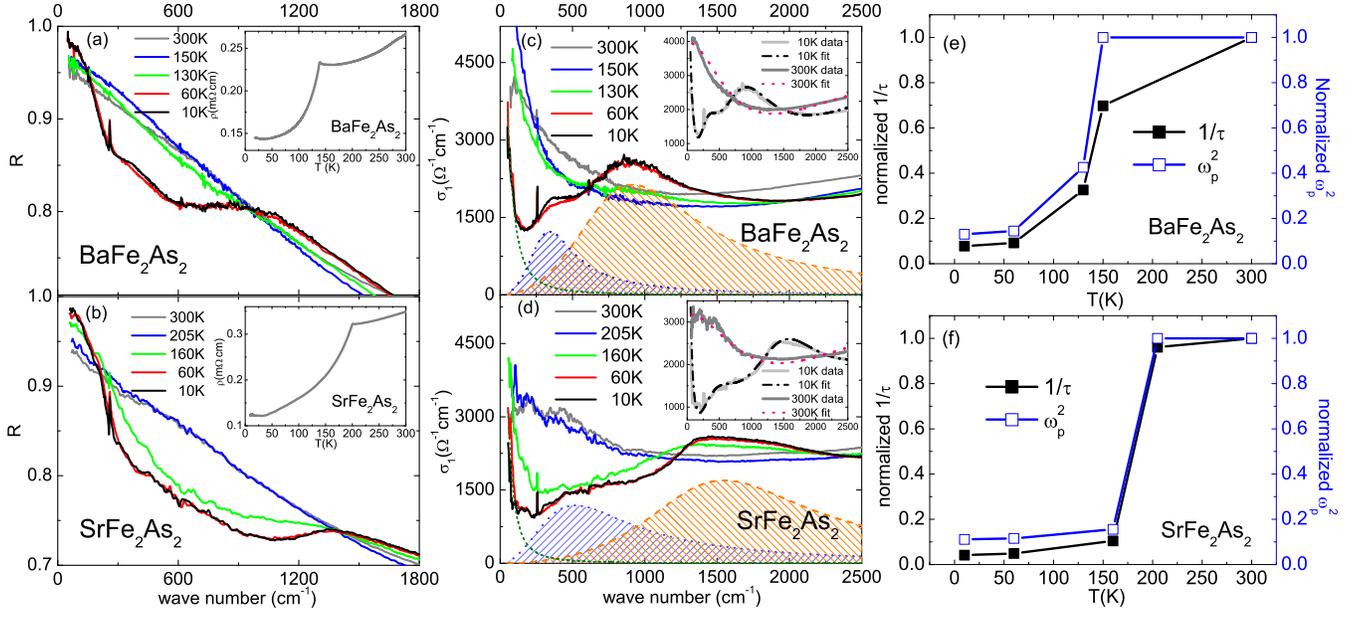}%
\vspace*{-0.20cm}%
\caption{\label{fig:R}(Color online) Left panel: R($\omega$) for
(a) BaFe$_2$As$_2$ and (b) SrFe$_2$As$_2$ below 1800 \cm. Inset:
the dc resistivity. Middle panel: $\sigma_1(\omega)$ for (c)
BaFe$_2$As$_2$ and (d) SrFe$_2$As$_2$ below 2500 \cm. The Drude
term (short dash green line) and the first two Lorentz peaks
abstracted from a Drude-Lorentz fit for T=10 K is shown at the
bottom. Inset: experimental data and the fit for T=10 and 300 K
below 2500 \cm. Right panel: normalized 1/$\tau$ (black) and
$\omega_p^2$ (blue) for (e) BaFe$_2$As$_2$ and (f)
SrFe$_2$As$_2$.}
\end{figure*}

Single crystals of BaFe$_2$As$_2$ and SrFe$_2$As$_2$ (space group
I4/mmm) were grown from the FeAs flux method.\cite{Chen3}  The
plate-like crystals could be easily cleaved, resulting in very
shinny surface. The T-dependent dc resistivity was measured by the
four contact technique in a Quantum Design PPMS. As shown in Fig.
1, both samples show metallic T-dependence in the whole
measurement temperature range. The resistivity drops sharply at
138 K and 200 K for BaFe$_2$As$_2$ and SrFe$_2$As$_2$,
respectively, which are ascribed to the formation of SDW order.

The optical reflectance measurements were performed on a
combination of Bruker IFS 66v/s, 113v and a grating-type
spectrometers on newly cleaved surfaces (ab-plane) for
AFe$_2$As$_2$ (A=Ba, Sr) single crystals in the frequency range
from 40 to 50000 cm$^{-1}$. An \textit{in situ} gold and aluminium
overcoating technique was used to get the reflectivity
R($\omega$). The real part of conductivity $\sigma_1(\omega)$ is
obtained by the Kramers-Kronig transformation of R($\omega$).

The main panels of Fig. 1 (a) and (b) focus on the low frequency
R($\omega$) up to 1800 \cm. For both compounds, R($\omega$)
exhibits a metallic response, and approaches to unity at zero
frequency. The most prominent feature is a substantial suppression
in R($\omega$) for \emph{T}$<$\emph{T}$_{SDW}$, which is a strong
optical evidence for the formation of energy gaps. The
low-$\omega$ reflectance increases faster towards unity at zero
frequency than those at high \emph{T}. As a consequence, one can
see a rather sharp low-$\omega$ reflectance edge. This indicates
clearly that the Fermi surfaces are only partially gapped and the
compounds are still metallic below \emph{T}$_{SDW}$. The change of
R($\omega$) from a overdamped linear-$\omega$ dependent behavior
to a well-defined reflectance edge upon cooling the sample into
SDW ordered state immediately suggests a dramatic reduction of the
carrier scattering rate, while its low-energy location implies a
considerable reduction of carrier density. A quantitative analysis
will be given below. It is noted that the low-\emph{T} R($\omega$)
displays an almost linear-$\omega$ dependence over a certain
frequency range below the suppression. This special shape of the
suppression leads to the two-peak structure in optical
conductivity.

The middle panels of Fig. 1 show the conductivity spectra
$\sigma_1(\omega$) below 2500 \cm. The Drude-like conductivity can
be observed for all spectra at low frequencies. For BaFe$_2$As$_2$
($T_{SDW}$=138 K), a weak feature around 890 \cm develops for
T=130 K in $\sigma_1(\omega$), then the spectra are severely
suppressed at low frequencies for 60 K and 10 K, resulting in a
pronounced double-peak character at 360 and 890 \cm. Associated
with the low-$\omega$ reflectance edge, a very sharp and narrow
Drude component emerges below the double peaks. Very similar
features can be seen for SrFe$_2$As$_2$ crystal, but the double
peak features appear at higher energies, \textit{i.e.} 500 and
1360 \cm, being consistent with the higher $T_{SDW}$ for
SrFe$_2$As$_2$. The electrodynamics of broken symmetry ground
states, such as the superconducting and density wave states, have
been well explored and understood. Due to different coherence
factors, a density wave state behaves different from an s-wave
superconductor at the gap frequencies in optical conductivity. In
an s-wave superconducting state at $T=0$, the absorption smoothly
rises at the gap frequency, while for a density wave state, a
sharp maximum appears in conductivity at the gap
frequency.\cite{Degiorgi,Vescoli} Based on those studies, we can
identify the double peak energies as the two SDW gaps. The
observation of two distinct SDW gaps should be associated with
different Fermi surfaces, and reflect the multi-band property in
FeAs based compound. From the gap values and SDW transition
temperatures, we obtained the ratio of
2$\Delta$/$k_BT_{SDW}\approx$3.5-3.6 for the smaller gap, and
9-9.6 for the larger gap for the two compounds. The smaller gap
coincides roughly with the gap value expected by the conventional
BCS relation, while the large one is very different. Similar
two-gap behaviors were also found in optical measurement for the
itinerant SDW metal Cr.\cite{Lind}.

Due to the presence of two different gap values in the SDW ordered
state, direct information on where the FS sheets are gapped is
highly desired. Very recently, two angle resolved photoemission
spectroscopy (ARPES) investigations on BaFe$_2$As$_2$ single
crystals were reported, however, they yielded completely different
results from optics. Yang \emph{et al.}\cite{Feng} reported a
complete absence of gap opening for all bands at the Fermi level.
Liu \emph{et al.}\cite{Kaminski} observed a small circular-shaped
hole pocket centered at $\Gamma$ and a large electron pocket at X
point below T$_{SDW}$ (at 100 K) on BaFe$_2$As$_2$ which matches
well with full-potential linearized plane wave calculations.
However, the gap opening is also absent in this work. Apparently
the available ARPES data are in sharp contrast to our results. We
emphasize that optical measurement is a bulk probe, while ARPES
measurement strongly depends on the surface quality. We also noted
that the large area of FS seen by Liu \emph{et al.} below
T$_{SDW}$ is incompatible to the quantum oscillation experiment
which reveals a rather small residual FS (occupying only 2$\%$ of
the Brillouin zone).\cite{Sebastian} More experiments are needed
to resolve the inconsistency.

It is well-known that the low-$\omega$ Drude component comes from
the itinerant carrier contribution. The Drude spectral weight
determines the $\omega_p^2$ ($\omega_p$ is the plasma frequency),
which is proportional to \emph{n/m$_{eff}$} (where \emph{n} is the
carrier density, \emph{m$_{eff}$} is the effective mass); while
its width reflects the carrier scattering rate 1/$\tau$. To
qualitatively analyze the T-evolution of the free-carrier
component, we fit the optical response by the Drude-Lorentz model
for the whole \emph{T} and frequency range.
\begin{equation}
\epsilon(\omega)=\epsilon_\infty-{{\omega_p^2}\over{\omega^2+i\omega/\tau}}+\sum_{i=1}^N{{S_i^2}\over{\omega_i^2-\omega^2-i\omega/\tau_i}}.
\label{chik}
\end{equation}
Here, $\epsilon_\infty$ is the dielectric constant at high energy,
the middle and last terms are the Drude and Lorentz components,
respectively. We use one Drude component for the free-carrier
response, and several Lorentz terms to fit the high frequency
$\sigma_1(\omega)$, including the double-peak SDW gap below
T$_{SDW}$, a pronounced mid-infrared feature near 5000 \cm (Fig. 2
(b)), and some Lorentz components above 16000 \cm for interband
transitions. The experimental $\sigma_1(\omega)$ data could be
well reproduced by the fit (see the insets of Fig. 1 (c) and (d)).

We are now mainly concerned with the evolution of the itinerant
carriers. For BaFe$_2$As$_2$, the plasma frequency
$\omega_p\approx$12900 \cm and scattering rate 1/$\tau\approx$700
\cm at 300 K reduce to 4660 \cm and 55 \cm at 10 K, respectively.
For SrFe$_2$As$_2$, $\omega_p\approx$13840 \cm and
1/$\tau\approx$950 \cm at 300 K reduce to 4750 \cm and 40 \cm at
10 K, respectively. Figure 1 (e) and (f) shows the variations of
1/$\tau$ and $\omega_p^2$ with temperature for the Drude term.
Both parameters are normalized to their 300 K value. Provided the
effective mass of itinerant carriers does not change with
temperature, then the residual carrier density is only 12$\%$ of
that at high temperature for both compounds. This means that
roughly 88$\%$ of FS is removed by the gapping associated with SDW
transitions. On the other hand, the scattering rate was reduced by
about 92-96$\%$. Therefore, the opening of the SDW partial gap
strongly reduces the scattering channel, leading to a metallic
behavior with enhanced dc conductivity in the gapped state.

The above observations have important implication for the driving
mechanism of SDW instability. As mentioned above, the key issue
here is whether an itinerant picture based on FS nesting or a
local superexchange mechanism is a proper approach. Our optical
studies clearly demonstrate that the parent compound has a high
itinerant carrier density with the plasma frequency a bit higher
than 1.5 eV before SDW transition, and is rather metallic both
above and below SDW ordering temperatures. Furthermore, the
partial gap openings below SDW ordering temperatures are
consistent with the expectation of nesting scenario where the
temperature dependence of the gap should resemble that of BCS
theory. On this basis, we think that the local picture is less
favored, and the itinerant scenario provides more reasonable
explanation for the driving mechanism.

\begin{figure*}
\includegraphics[width=7in]{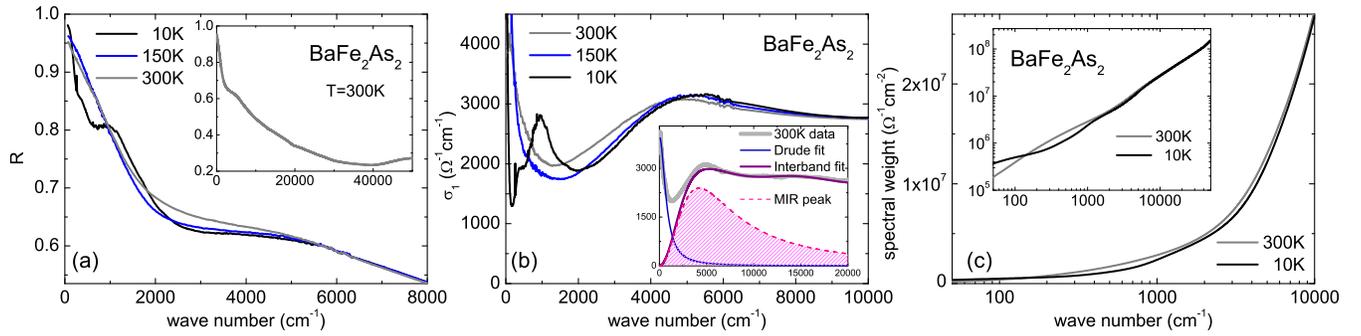}%
\vspace*{-0.20cm}%
\caption{\label{fig:S1}(Color online) (a) R($\omega$) for
BaFe$_2$As$_2$ below 8000 \cm. Inset: R($\omega$) below 50000 \cm.
(b) $\sigma_1(\omega)$ below 10000 \cm (together with the
low-$\omega$ extrapolation based on the dc conductivity). Inset:
$\sigma_1(\omega)$ below 20000 \cm (T=300 K), the intra- and
interband terms, and the mid-infrared peak from the Drude-Lorentz
fit. (c) The spectral weight below 10000 \cm. Inset: the spectral
weight up to 50000 \cm.}
\end{figure*}

Besides the dramatic spectral change at low frequencies, both
compounds display very similar and pronounced spectral feature at
the mid-infrared region. Take BaFe$_2$As$_2$ as an example, the
mid-infrared component takes up a large spectral weight as shown
in the inset of Fig. 2 (b). The peak at such a high energy is
usually ascribed to the interband transition. However, a puzzling
problem is that the spectra \emph{below} the peak energy exhibit
an apparent T-dependence that R($\omega$) is obviously suppressed
with decreasing T below 5000 \cm (Fig. 1 (a)). Such a gap-like
feature is present at all temperatures. Indeed, an analysis for
the spectral weight in $\sigma_1(\omega)$ revealed that the
suppressed spectral weight below 5000 \cm is transferred to higher
energies. Figure 2 (c) plots the spectral weight for
$\sigma_1(\omega)$ at 10 and 300 K. Below 200 \cm, the narrowing
Drude peak with decreasing T (see the extrapolation data in Fig. 2
(b)) yields a growing dc conductivity thus a larger low frequency
spectral weight for T=10 K. Above 200 \cm, the SDW double-gap
develops which strongly reduces the low T Drude weight, leading to
the first suppression below 1000 \cm in Fig. 2 (c). The lost Drude
weight fills into the SDW double-peak, and the total spectral
weight is almost recovered around 2000 \cm for 10 K. Then the
T-dependent suppression before the mid-infrared peak results in
the second spectral weight suppression at 10 K near 3000 \cm. The
lost weight finally recovers at about 8000 \cm. We would like to
emphasize that this gap-like feature is not directly related to
the SDW order. This is because (1) the mid-infrared suppression
feature is present above T$_{SDW}$, (2) similar features exist for
K- or Co-doped superconducting samples where the SDW order and the
associated low energy gap structures are completely
absent,\cite{Li} and (3) the energy scale is much larger than the
SDW gaps. Usually a gap formation is associated with a broken
symmetry state. As the AFe$_2$As$_2$ compounds are in their
paramagnetic phase with a tetragonal crystal structure above
T$_{SDW}$, both magnetic and crystal structures are in very high
symmetry state, one would hardly expect an even higher magnetic or
crystal structural symmetry at higher temperatures. Definitely,
further experimental and theoretical works are necessary to
understand this high energy gap-like behavior.

To summarize, the ab-plane optical measurements of AFe$_2$As$_2$
(A=Ba, Sr) single crystals were performed. For the SDW state, our
findings indicate the Fermi surface is largely affected by the SDW
transition. Based on a Drude-Lorentz model, we estimate that about
88$\%$ itinerant carriers were removed by the gapping of Fermi
surfaces. Meanwhile, the carrier scattering rate was reduced by
92-96$\%$. More importantly, we find two distinct gaps in the SDW
ordered state in AFe$_2$As$_2$. For the high frequency regions,
R($\omega$) show anomalous T dependence even above T$_{SDW}$ near
the 5000 \cm mid-infrared component, which suggests the role of Fe
3\emph{d} electrons in AFe$_2$As$_2$ system might be very complex
and unusual. Since AFe$_2$As$_2$ (A=Ba, Sr) systems have rather
high conducting carrier density at high T, a driving mechanism
based on an itinerant picture for the SDW instability is thus
favored.

\begin{acknowledgments}
This work is supported by the National Science Foundation of
China, the Knowledge Innovation Project of the Chinese Academy of
Sciences, and the 973 project of the Ministry of Science and
Technology of China.
\end{acknowledgments}


\end{document}